\documentclass[sigconf]{acmart}
\usepackage{balance}

\usepackage{amsmath,amsfonts}
\usepackage{algorithm}
\usepackage{algpseudocode}
\usepackage{wrapfig, blindtext}
\usepackage[export]{adjustbox}
\usepackage{graphicx}
\usepackage{textcomp}

\usepackage[utf8]{inputenc} 
\usepackage[T1]{fontenc}    
\usepackage{hyperref}       
\usepackage{url}            
\usepackage{booktabs}       
\usepackage{amsfonts}       
\usepackage{nicefrac}       
\usepackage{microtype}      
\usepackage{xcolor}         
\usepackage{colortbl}

\usepackage{diagbox}
\usepackage{mathtools}
\usepackage{pifont}
\usepackage{multirow}
\usepackage{multicol}
\usepackage{threeparttable}
\usepackage{comment}
\usepackage{subfigure}
\usepackage{subcaption}

\usepackage{enumitem}
\usepackage{tikz}
\usepackage{pgfplots}
\usepackage[multiple]{footmisc}
\usepackage{bm}
\usepackage{booktabs}
\usepackage{array}
\usepackage{siunitx}
\usepackage{tabularx}
\usepackage{geometry}
\usepackage{lipsum}

\usepackage{contour}
\usepackage{ulem}[normalem]
\usepackage{dsfont}
\usepackage{makecell}

\usepackage{float}

\newcommand\TODO[1][]{{\color{orange}[TODO\ifthenelse{\equal{#1}{}}{}{: #1}]}}
\newcommand\SEC\section
\newcommand\SSEC\subsection
\newcommand\SSSEC\subsubsection

\newcommand{\rom}[1]{\uppercase\expandafter{\romannumeral #1\relax}}

\AtBeginDocument{%
  \providecommand\BibTeX{{%
    \normalfont B\kern-0.5em{\scshape i\kern-0.25em b}\kern-0.8em\TeX}}}

\setcopyright{acmlicensed}
\copyrightyear{2025}
\acmYear{2025}
\acmDOI{ }

\acmConference[CIKM '25]{Proceedings of the 34th ACM International Conference on Information and Knowledge Management}{November 10--14,
2025}{Seoul, Korea}
\acmBooktitle{Proceedings of the 34th ACM International Conference on Information and Knowledge Management (CIKM '25), November 10--14,
2025, Seoul, Korea}

\begin{document}

\title{Continual Recommender Systems}

\author{Hyunsik Yoo}
\email{hy40@illinois.edu}
\affiliation{
  \institution{University of Illinois Urbana-Champaign}
  \city{}
  \state{}
  \country{Urbana, IL, USA}
}

\author{SeongKu Kang}
\email{seongkukang@korea.ac.kr}
\affiliation{
  \institution{Korea University}
  \city{}
  \state{}
  \country{Seoul, Republic of Korea}
}

\author{Hanghang Tong}
\email{htong@illinois.edu}
\affiliation{
  \institution{University of Illinois Urbana-Champaign}
    \city{}
  \state{}
  \country{Urbana, IL, USA}
}

\renewcommand{\shortauthors}{Hyunsik Yoo, SeongKu Kang, \& Hanghang Tong}

\ccsdesc[500]{Information systems~Data mining}
\ccsdesc[500]{Computing methodologies~Machine learning}

\keywords{Recommender systems; Continual learning; Personalization}

\begin{abstract}
Modern recommender systems operate in uniquely dynamic settings: user interests, item pools, and popularity trends shift continuously, and models must adapt in real time without forgetting past preferences. While existing tutorials on continual or lifelong learning cover broad machine learning domains (e.g., vision and graphs), they do not address recommendation-specific demands—such as balancing stability and plasticity per user, handling cold-start items, and optimizing recommendation metrics under streaming feedback. This tutorial aims to make a timely contribution by filling that gap. We begin by reviewing the background and problem settings, followed by a comprehensive overview of existing approaches. We then highlight recent efforts to apply continual learning to practical deployment environments, such as resource-constrained systems and sequential interaction settings. Finally, we discuss open challenges and future research directions. We expect this tutorial to benefit researchers and practitioners in recommender systems, data mining, AI, and information retrieval across academia and industry.

\end{abstract}




\maketitle

\section{Introduction}

Modern recommender systems operate in dynamic environments where user preferences, behaviors, and available items constantly change. 
As new data continually flows in, models can quickly become stale and outdated \cite{wang2023structure}.
To maintain high recommendation quality, models must adapt to evolving user preferences and accommodate new users and items.
Traditional retraining on the full dataset is often too slow and resource-intensive to keep pace with rapid data updates. 
A more efficient alternative is fine-tuning, which incrementally updates models using only new data. 
For example, a model pre-trained on one year's data can be fine-tuned weekly with the latest data.

Continual learning builds on fine-tuning as a foundational strategy for processing new data effectively.
A key challenge is balancing plasticity (adapting to new knowledge) and stability (preserving prior knowledge) \cite{yoo2025embracing}. 
Moreover, this balance may need to be personalized for users with different behaviors, which is a core philosophy in recommendation modeling. 
Naïve fine-tuning can overly favor plasticity, leading to catastrophic forgetting, where new learning overwrites previously acquired knowledge.
Failing to preserve prior knowledge can severely degrade recommendation quality and lead to inconsistent user experiences, particularly for returning users or in long-tail item scenarios.
Furthermore, recommender systems operate in a variety of environments that present practical challenges for continual updates.
One such example is resource-constrained environments, where systems must cope with limited memory, computational capacity, or energy budgets—such as in edge devices or latency-sensitive applications.
Deploying continual learning models under these constraints introduces additional difficulties in ensuring real-time, adaptive updates \cite{lee2024continual}.


Given the growing research attention and practical needs surrounding continual learning for recommender systems, our tutorial aims to make a timely contribution by providing a systematic review of prior work and outlining future directions.
We begin with an introduction to the background, problem settings, and applications of continual learning in recommender systems (Part~\rom{1}). 
Next, we provide a comprehensive overview of existing methods, organized into two main categories: experience replay-based strategies (Part~\rom{2}) and regularization-based strategies (Part~\rom{3}). 
We then discuss approaches that go beyond traditional setups, focusing on resource-constrained environments and sequential interaction environments (Part~\rom{4}). 
Finally, we conclude with a discussion of open challenges and future directions (Part~\rom{5}).
The website of this tutorial is available \href{https://www.idea.korea.ac.kr/research/tutorial-continual-recommender-systems}{\textcolor{blue}{here}}.
\section{Target Audience and Benefits}


This tutorial is intended for researchers and practitioners in data mining, artificial intelligence, information retrieval, and related fields. 
The audience is expected to have basic knowledge of probability, linear algebra, and machine learning, but no prior familiarity with specific continual learning algorithms is required. 

\textbf{This tutorial is relevant and valuable to the CIKM community due to the practicality and generalizability of both ``recommender systems'' and ``continual learning''.}

\textbf{(1) Recommender systems:} Many online platforms influence what people hear, watch, buy, or interact with. 
Recommender systems are widely integrated into these platforms to enhance user satisfaction, directly impacting daily user experiences. 
Improving recommendation quality is crucial not only for user engagement but also for driving business revenue in many applications.
Moreover, personalization techniques discussed in this tutorial are applicable to other domains that require user modeling, such as information retrieval, graph mining, and natural language processing.

\textbf{(2) Continual learning:} Real-world decision-making systems, including recommendation models, must adapt to continuously arriving data. 
While periodic retraining is common, more frequent updates are often infeasible due to resource constraints and the scale of industrial systems. 
Continual learning enables efficient and effective fine-tuning on new data, enhancing both computational efficiency and adaptability to evolving user preferences, as well as to new users and items. 
Our tutorial provides a comprehensive overview of prior efforts and offers a friendly walk-through to support audience understanding.


\section{Outline and Timeline}
The length of the tutorial will be half day, i.e., 3 hours plus breaks.
\begin{itemize}[leftmargin=*] 
    \item Part~\rom{1}: Introduction and Background (30 min)
    \begin{itemize}
        \item Problem definitions and settings
        \item Key challenges 
        \item Applications and practical use cases
    \end{itemize}
    \item Part~\rom{2}: Experience-Replay-based Methods (35 min)
    \begin{itemize}
        \item Sample selection for experience replay
        \item Replay-based model enhancement
    \end{itemize}
    \item Part~\rom{3}: Regularization-based Methods (35 min)
    \begin{itemize}
        \item What knowledge to regularize
        \item Which temporal knowledge to regularize
        \item Personalization of regularization
    \end{itemize}
    \item Short Break (5 min)
    \item Part~\rom{4}: Beyond Traditional Settings (35 min)
    \begin{itemize}
        \item Resource-constrained environments
        \item Sequential interaction environments
    \end{itemize}
    \item Part~\rom{5}: Open Challenges and Future Directions (35 min)
    \begin{itemize}
        \item Trustworthiness (e.g., fairness, explainability, robustness)
        \item Adaptation to foundational models
        \item Unified models for recommendation and search
    \end{itemize}
    \item Conclusion (5 min)
\end{itemize}
A list of the most important references that will be covered in the tutorial is provided \href{https://www.idea.korea.ac.kr/research/tutorial-continual-recommender-systems}{\textcolor{blue}{here}}.

\section{Tutorial Content}

\subsection{Part~\rom{1}: Introduction and Background}
We begin by discussing the \textbf{background and motivations} for studying continual learning in recommender systems. Next, we provide an overview of existing problem settings for training and inference. We highlight \textbf{key challenges} such as mitigating forgetting, adapting to changing user preferences and item popularity, handling new users and items, and personalizing the stability–plasticity balance. Finally, we present \textbf{applications} and practical use cases.



\subsection{Part~\rom{2}: Experience-Replay-based Methods}
Experience replay is one of the most widely used strategies, which involves reusing past user interactions.
Two key aspects are critical in this approach: which past experiences to retain and how to effectively leverage them to enhance the model. 
This part introduces representative methods that address these two aspects.
\begin{itemize}[leftmargin=*] 
    \item \textbf{Sample Selection for Replay:} Since storing all historical data is infeasible in dynamic environments, selecting which samples to retain is a key challenge.
    Some methods prioritize samples based on interaction frequency \cite{mi2020ader, ahrabian2021structure}, while others employ error-driven selection, maintaining samples that the model previously mispredicted \cite{cai2022reloop, zhu2023reloop2}. 
    There have also been attempts to identify influential samples by analyzing their impact on the model’s training behavior \cite{zhang2024influential}, or to store user-item-context features using a dedicated encoder in a parametric space \cite{qin2025d2k}.
    We review recent methods for constructing an informative knowledge buffer.
    
    \item \textbf{Replay-based Model Enhancement:} Once past samples are selected and stored, a central question is how to integrate them effectively during model training and prediction.
    One line of work introduces specialized loss functions that minimize prediction errors in comparison to earlier model versions \cite{cai2022reloop}. 
    Another approach estimates and compensates for prediction errors using the outputs of previous models stored in the buffer \cite{zhu2023reloop2}.
    A more recent method injects the knowledge stored in the parametric space into the model via a dedicated adaptation module \cite{qin2025d2k}.
    We review recent methods that explore these replay-based model enhancement strategies.
\end{itemize}

\subsection{Part~\rom{3}: Regularization-based Methods}
Regularization-based methods constrain the current model's parameters using those learned at different time steps. Early approaches focused on improving stability by imposing constraints between past and current parameters to mitigate forgetting. Methods vary in the type of knowledge used for regularization (e.g., what is regularized and from which time point—past or future relative to current training) and whether the approach is personalized to individual users, as outlined below:

\begin{itemize}[leftmargin=*] 
    \item \textbf{What Knowledge to Regularize:}  
    User representations are typically the primary focus for regularization, aligning current representations with their historical counterparts.\footnote{Item-side regularization can also be applied analogously.} 
    Methods differ in how they define this knowledge. One approach incorporates local structure (i.e., a user’s neighboring items), global structure (i.e., the distribution of distances between the user and item clusters), and self-information from the interaction network~\cite{xu2020graphsail, wang2023structure}.  
    Another line of work maximizes mutual information (e.g., using InfoNCE) between past and current layer-wise parameters of GNN-based models~\cite{wang2021graph, yoo2025embracing}.

    \item \textbf{Which Temporal Knowledge to Regularize:}  
    Most methods regularize current parameters using previously learned ones (backward knowledge) to enhance stability~\cite{xu2020graphsail, wang2023structure, wang2021graph}. Recent work incorporates forward knowledge—such as parameters obtained through fine-tuning on current data—to explicitly promote plasticity when adapting to new information is critical~\cite{yoo2025embracing}. By selectively applying each type of knowledge based on context, bidirectional regularization enables a more effective balance between stability and plasticity.

    \item \textbf{Personalized Regularization:}  
    Because recommendation is inherently user-centered, personalizing the stability–plasticity trade-off is crucial. Dynamic users with shifting preferences benefit from regularization that encourages plasticity, while stable users benefit from regularization that reinforces stability. One method learns personalized regularization weights through a multi-layer perceptron that captures user preference shifts, modeled as changes in distance to item cluster centers~\cite{wang2023structure}. Another approach uses deterministic weights, selectively applied based on user dynamics (e.g., high or low preference shifts), to prioritize either plasticity or stability~\cite{yoo2025embracing}.

\end{itemize}

\subsection{Part~\rom{4}: Beyond Traditional Settings}
Real-world deployments introduce additional challenges that go beyond traditional settings. 
In this part, we focus on two important practical environments, both of which pose unique challenges for continual adaptation and efficient learning.
\begin{itemize}[leftmargin=*] 
    \item \textbf{Resource-Constrained Environments}: Many recommender systems operate under constraints such as limited memory, computational capacity, or energy budgets—particularly in edge devices or latency-sensitive applications \cite{kang2023distillation}.
    These constraints limit the feasibility of storing extensive past histories or retaining previously trained large models.
    We review recent approaches for selective experience replay and the adaptation of lightweight models via knowledge distillation, with the goal of reducing computational overhead while maintaining adaptability over time~\cite{lee2024continual}. 
    \item \textbf{Sequential Interaction Environments}: User interactions are inherently sequential in many applications, ranging from session-based settings to long-term user engagement in streaming platforms. 
    However, most prior continual learning studies have focused on conventional models (e.g., matrix factorization), while dedicated approaches for sequential pattern models (e.g., SASRec \cite{sasrec}) remain underexplored.
    We review the unique challenges of directly applying existing methods to sequential models, as well as recent efforts to address these challenges \cite{linrec, lee2025leveraging}.

\end{itemize}

\subsection{Part~\rom{5}: Future Directions}
In this part, we discuss potential future directions for continual recommender systems, including improving trustworthiness, enabling continual learning in large foundational models, and unifying search and recommendation within a single framework.

\begin{itemize}[leftmargin=*] 
    \item \textbf{Trustworthiness:}
    Fairness is a key aspect of trustworthiness and is especially critical in dynamic environments, where it can deteriorate over time due to factors such as negative feedback loops or growing data imbalance across user or item groups. Recommender systems also involve multiple fairness dimensions, including user-side, item-side, and two-sided fairness. While there has been progress on user- and item-side fairness in continual settings~\cite{yoo2024ensuring, morik2020controlling, guo2025enhancing, ge2021towards, zhu2021popularity}, two-sided fairness remains relatively underexplored~\cite{tang2025model}. Moreover, maintaining fairness remains practically challenging due to evolving user behavior, and inherent trade-offs with accuracy. Other aspects of trustworthiness—such as adversarial robustness, diversity, and explainability—also warrant further attention in continual learning contexts.

    \item \textbf{Adaptation to Foundational Models:}
    Foundational models, such as large transformer-based architectures, are increasingly used in recommendation tasks, including LLM-based generative recommenders often fine-tuned via LoRA. Although recent work has explored continual variants of LoRA, these efforts are either focused on other domains (e.g., vision) or overlook personalization in the stability–plasticity trade-off~\cite{wu2025s, shi2024preliminary}. Further research is needed to adapt continual learning techniques effectively to such large models in recommender systems.

    \item \textbf{Unified Models for Recommendation and Search:}
    Recently, there have been a few attempts to unify search and recommendation into a single framework \cite{penha2024bridging, shi2025unified}.
    Accordingly, new challenges emerge for continual learning beyond those found in each domain individually.
    While search and recommendation each exhibit domain-specific dynamics—such as query drift in search and preference shifts in recommendation—the unified setting introduces additional complexity. 
    Models must learn from heterogeneous signals (e.g., clicks, queries, purchases) and maintain consistency across distinct objectives and feedback modalities. 
    Moreover, knowledge transfer between the two tasks must be handled carefully to avoid negative interference, especially under distributional shifts. 
    Designing continual learning strategies that can effectively balance shared and task-specific knowledge is critical for unified systems.
    
\end{itemize}
\vspace{-2mm}
\section{Related Tutorials}
The following is a list of relevant tutorials that have been or will be presented at other prominent data mining and machine learning conferences, along with a brief discussion of their connections and differences compared to ours.
\vspace{-\topsep}
\begin{itemize}[leftmargin=*] 
    \item \textbf{Continual Learning with Deep Architectures (ICML ’21) 
    \item Lifelong Learning Machines (NeurIPS ’22)
    \item Never-Ending Learning, Lifelong Learning and Continual Learning (AAAI ’23)}
    \begin{itemize}
        \item \textbf{Connection:} These three tutorials cover the fundamentals of continual and lifelong learning in general machine learning. They introduce core techniques such as experience replay, regularization for mitigating forgetting, parameter isolation strategies, and dynamic architectures, and they discuss broad paradigms and benchmarks that support continual learning.
        \item \textbf{Difference:} They are general-purpose and focus on classic ML domains (e.g., vision, RL) and theory, without targeting recommender systems. As a result, they do not address recommendation-specific challenges such as evolving user–item interactions, personalization strategies, or user preference adaptation.
    \end{itemize}

  \item \textbf{Lifelong Learning for Cross-domain Recommender Systems (WWW ’23)}
    \begin{itemize}
      \item \textbf{Connection:} This tutorial introduces ``3C'' principle (Complement, Composite, Context) for continuously integrating knowledge of multiple domains in recommender systems.
      \item \textbf{Difference:} It primarily focuses on multi-domain fusion aspect, whereas our tutorial covers real-time, single-domain recommendation updates and personalization.
    \end{itemize}
    
    \item \textbf{Continual Graph Learning (SDM ’23; WWW ’23; AAAI ’24)}
    \begin{itemize}
      \item \textbf{Connection:} This tutorial covers continual learning on various types of dynamic graphs, using techniques such as replay, regularization, and dynamic benchmarks.
      \item \textbf{Difference:} 
      It focuses on general graph-ML frameworks, whereas our tutorial drills into recommendation-specific goals like adapting to user preference shifts, item popularity changes, and personalized stability–plasticity trade-offs.
    \end{itemize}

\end{itemize}

\section{Presenter Biography}
The presenters and contributors of this tutorial are Hyunsik Yoo, SeongKu Kang, and Hanghang Tong. Their biographies and areas of expertise are provided below.

\textbf{Hyunsik Yoo} is a fourth-year Ph.D. student in the Siebel School of Computing and Data Science at the University of Illinois Urbana-Champaign. His research focuses on developing data mining and machine learning techniques for recommender systems and graph mining models that are adaptive, trustworthy, and user-inclusive. His work has been published in major conferences, including KDD, SIGIR, TheWebConf, WSDM, and ICML. He has also served as a program committee member or reviewer for venues such as KDD, CIKM, TheWebConf Companion, AAAI, NeurIPS, DSAA, and TIST.

\textbf{SeongKu Kang} is an Assistant Professor in the Department of Computer Science and Engineering at Korea University.
Prior to that, he was a postdoctoral researcher at the University of Illinois Urbana-Champaign.
His research interests lie in data mining, recommender systems, and information retrieval.
He has published more than 30 papers in major conferences such as KDD, TheWebConf, CIKM, SIGIR, and EMNLP.
He has also actively contributed to the research community by serving as a PC member or reviewer for venues including KDD, TheWebConf, AAAI, SIGIR, ACL, SDM, TIST, and TKDE, and was recognized as an outstanding reviewer~at~KDD.

\textbf{Hanghang Tong} is a professor in the Siebel School of Computing and Data Science at the University of Illinois Urbana-Champaign. Before that he was an associate professor at Arizona State University. He received his M.Sc. and Ph.D. degrees from Carnegie Mellon University in 2008 and 2009, both in Machine Learning. His research interest is in large scale data mining for graphs and multimedia. He has received several awards, including  IEEE ICDM Tao Li award (2019), SDM/IBM Early Career Data Mining Research award (2018), NSF CAREER award (2017), ICDM 10-Year Highest Impact Paper awards (2015 and 2022), and several best paper awards. He has published over 300 refereed articles. He was the Editor-in-Chief of ACM SIGKDD Explorations (2018-2022) and is an associate editor of ACM Computing Surveys. He is a distinguished member of ACM, a fellow of IEEE, and a university scholar.

\vspace{2mm}
\noindent\textbf{GenAI Usage.}
OpenAI GPT-4 was used throughout the manuscript for grammar and clarity. All AI-generated suggestions were manually reviewed and approved by the authors. 


\normalem
\bibliographystyle{ACM-Reference-Format}
\balance
\bibliography{references}

\end{document}